\documentclass[aps,pccp,reprint,superscriptaddress]{revtex4-1}
\usepackage{graphicx} 
\usepackage{amsmath} 
\usepackage{amsfonts} 
\usepackage{amssymb} 
\usepackage{threeparttable} 
\usepackage{mhchem}
\usepackage{appendix}

\usepackage{epstopdf}

\begin{document}

\title{Assessment of density functional approximations for the hemibonded structure of water dimer radical cation}

\date{\today}

\author{Piin-Ruey Pan}
\affiliation{Department of Physics, National Taiwan University, Taipei 10617, Taiwan}

\author{You-Sheng Lin}
\affiliation{Department of Physics, National Taiwan University, Taipei 10617, Taiwan}

\author{Ming-Kang Tsai}
\affiliation{Department of Chemistry, National Taiwan Normal University, Taipei 11677, Taiwan}

\author{Jer-Lai Kuo}
\email{jlkuo@pub.iams.sinica.edu.tw.}
\affiliation{Institute of Atomic and Molecular Sciences, Academia Sinica, Taipei 10617, Taiwan}

\author{Jeng-Da Chai}
\email{jdchai@phys.ntu.edu.tw.}
\affiliation{Department of Physics, National Taiwan University, Taipei 10617, Taiwan}
\affiliation{Center for Theoretical Sciences and Center for Quantum Science and Engineering, National Taiwan University, Taipei 10617, Taiwan}

\begin{abstract}

Due to the severe self-interaction errors associated with some density functional approximations, conventional density functionals often fail to dissociate the hemibonded structure of water dimer radical cation 
(H$_2$O)$_2$$^+$ into the correct fragments: H$_2$O and H$_2$O$^+$. Consequently, the binding energy of the hemibonded structure (H$_2$O)$_2$$^+$ is not well-defined. For a comprehensive comparison 
of different functionals for this system, we propose three criteria: (i) The binding energies, (ii) the relative energies between the conformers of the water dimer radical cation, and (iii) the dissociation curves 
predicted by different functionals. 
The long-range corrected (LC) double-hybrid functional, $\omega\mbox{B97X-2(LP)}$ [J.-D. Chai and M. Head-Gordon, \textit{J. Chem. Phys.}, 2009, \textbf{131}, 174105.], is shown to perform reasonably well 
based on these three criteria. Reasons that LC hybrid functionals generally work better than conventional density functionals for hemibonded systems are also explained in this work. 

\end{abstract}

\maketitle

\section{Introduction}

Water can be decomposed when it is exposed to high-energy flux. The products of water radiolysis may contain various radical species, \textit{e.g.}~hydrogen atoms (H), hydroxide radicals (OH), oxygen anions (O$^-$), and water 
cations (H$_2$O$^+$), depending on the radiation infrastructure setup. For example the overall decomposition scheme activated by $\beta$ particles has been outlined by Garrett \textit{et al.}~in 2005 \cite{Garrett} where three main channels of decomposition were listed. 
The cationic channel leads to the formation of ionized water living in about several tens femtoseconds and hydrated electron, followed by the generation of hydronium (H$_3$O$^+$) and OH radicals through proton 
transfer process \cite{LaVerne, Furuhama}. The energized-neutral and anionic channels could result in the cleavage the oxygen-hydrogen chemical bonds to produce hydrogen and oxygen derivatives, \textit{i.e.}~H, H$^-$, H$_2$, O, O$^-$, 
OH$^-$ \textit{etc}. Subsequent chemical reactions can progress further up to the desorption of stable gas molecules H$_2$ and O$_2$ being driven by those reactive radical species \cite{Garrett}. The cationic channel 
is therefore particularly interesting due to its dominant products --- OH radicals and solvated electrons. 

The smallest system to understand the chemical dynamics of ionized water is the water dimer radical cation (H$_2$O)$_2$$^+$, and it has been approached by several experimental studies in the past. Angel and 
Stace reported the predominant H$_3$O$^+$--OH central core by the collision-induced fragmentation experiment \cite{Angel} against the earlier theoretical assignment of charge-resonance hydrazine 
structure \cite{Barnett}. Dong \textit{et al.}~observed a weak signal corresponding the formation of (H$_2$O)$_2$$^+$ near the low-mass side of (H$_2$O)$_2$H$^+$ using 26.5 eV soft X-ray laser \cite{Dong}. 
Gardenier, Johnson, and McCoy reported the argon-tagged predissociation infrared spectra of (H$_2$O)$_2$$^+$ and assigned its structural pattern as a charge-localized H$_3$O$^+$--OH complex \cite{Gardenier}. 
Recently, Fujii's group reported the infrared spectroscopic observations of larger (H$_2$O)$_n$$^+$ clusters, $n = 3-11$ \cite{Fujii}, where the OH radical vibrational signal was clearly identified 
for $n\leqq5$ clusters, but vibrational signature of OH radical become inseparable due to the overlap with H-bonded OH stretch in $n>6$ . As being evidenced in the earlier studies \cite{Gardenier, Fujii}, theoretical investigations such as \textit{ab initio} electronic structure theory and density functional theory (DFT) play an important role in 
understanding the infrared spectroscopic features of the ionized water clusters. Because high-level \textit{ab initio} calculations are computationally prohibited for larger ionized water clusters, \textit{e.g.}~fully solvated 
cationic moiety, a reliable DFT method is necessary. 

\begin{figure}[h]
(a)
\includegraphics[scale=0.15,angle=90]{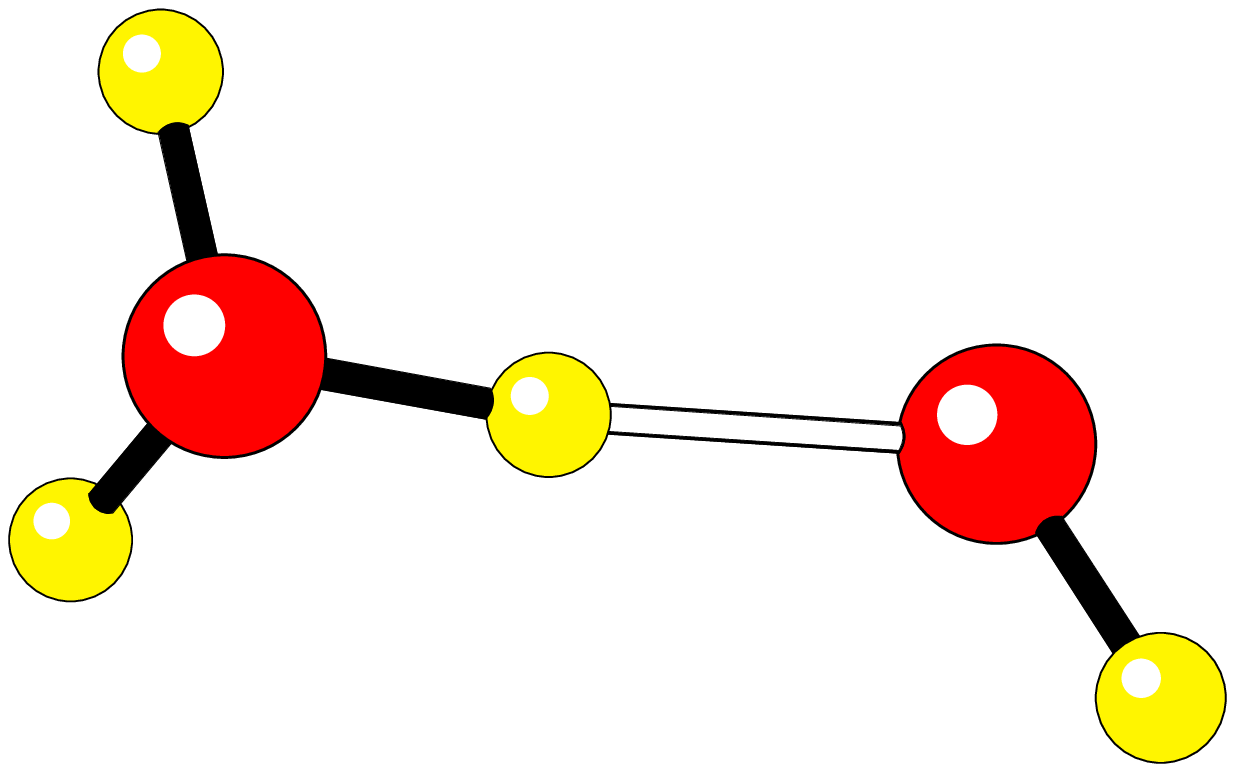}
(b)
\includegraphics[scale=0.15,angle=90]{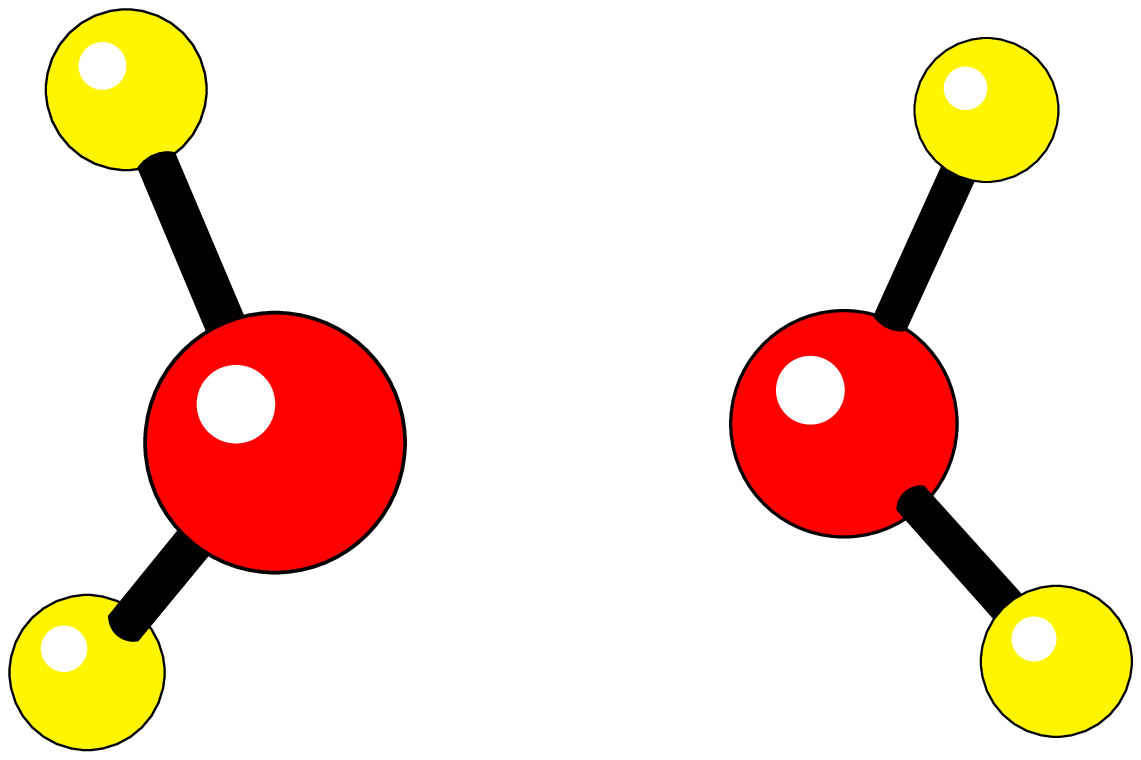}
(c)
\includegraphics[scale=0.15,angle=90]{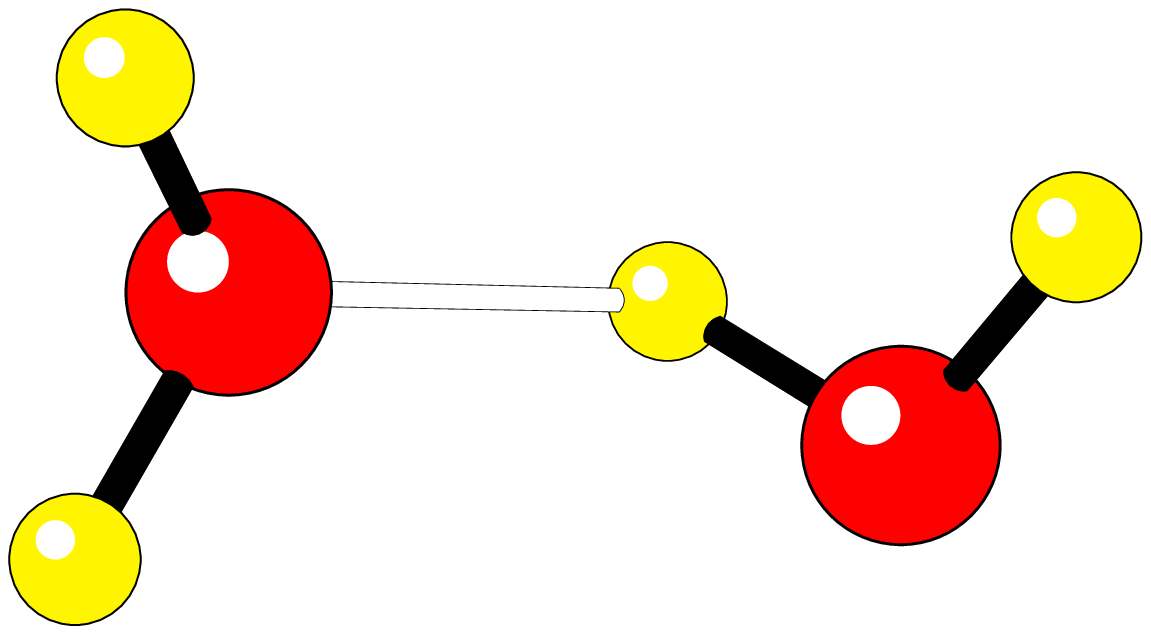}
\caption{(a) The proton transferred structure, (b) the hemibonded structure, and (c) the transition state between the structures of (a) and (b).}
\label{Fig:dimer}
\end{figure}

In earlier theoretical reports, two minimum structures of the water dimer radical cation have been identified: the proton transferred structure and the hemibonded structure, as shown in Fig.~\ref{Fig:dimer} \cite{JPCA1999,PCCP,Kim,Schaefer}. The previous DFT calculations have shown that many exchange-correlation (XC) functionals fail to predict reasonable results \cite{JPCA1999,PCCP,Kim} given rise to the presence of hemibonding interaction.
The hemibonding interaction, could be theoretically located in (H$_2$O)$_n$$^+$ systems, is notorious for the serious self-interaction errors (SIEs) associated with some density functional 
approximations. Both local density approximation (LDA) and generalized gradient approximations (GGAs) were reported to contain non-negligible amount of SIEs for describing the hemibonded structure \cite{JPCA1999, PCCP}. It has been suggested to adopt hybrid functionals with larger fractions of the exact Hartree-Fock (HF) exchange for more accurate results in the hemibonded 
structure \cite{JPCA1999,PCCP}. However, as the SIEs of functionals become larger at the dissociation limit, these suggested functionals can yield spurious barriers on their dissociation curves \cite{PCCP}, 
which can lead to unphysical results in molecular dynamics simulations. 

Clearly, more stringent criteria for choosing suitable functionals are needed. In this work, we propose three different criteria for a comprehensive 
comparison of different functionals for this system.

\section{Computational Methods}

Calculations are performed on the optimized geometries of the two structures of the water dimer cation and the transition state between them, optimized with the {\it ab initio} MP2 theory \cite{MP1934} and various 
XC functionals involving BLYP \cite{B88,LYP}, PBE \cite{PBE}, and M06L \cite{M06L}, which are pure density functionals (\textit{i.e}.~the fraction of HF exchange $\alpha_{\mathrm{HF}}=0.00$), 
B97 \cite{B97} with $\alpha_{\mathrm{HF}}=0.19$, B3LYP \cite{B88,LYP,B3LYP} with $\alpha_{\mathrm{HF}}=0.20$, PBE0 \cite{PBE0} with $\alpha_{\mathrm{HF}}=0.25$, 
M06 \cite{M06M062X} with $\alpha_{\mathrm{HF}}=0.27$, M05 \cite{M05} with $\alpha_{\mathrm{HF}}=0.28$, BH\&HLYP \cite{BHHLYP} with $\alpha_{\mathrm{HF}}=0.50$, 
M06-2X \cite{M06M062X} with $\alpha_{\mathrm{HF}}=0.54$, M05-2X \cite{M052X} with $\alpha_{\mathrm{HF}}=0.56$, M06HF \cite{M06HF} with $\alpha_{\mathrm{HF}}=1.00$, 
the $\omega\mbox{B97}$ series ($\omega\mbox{B97}$ \cite{omegaB97}, $\omega\mbox{B97}$X \cite{omegaB97}, $\omega\mbox{B97}$X-D \cite{omegaB97XD}, and $\omega\mbox{B97}$X-2(LP) \cite{omegaB97DH}), 
which are long-range corrected (LC) hybrid functionals (\textit{i.e}.~the fraction of HF exchange depends on the interelectronic distance \cite{optimal}), and the double-hybrid functional B2PLYP \cite{B2PLYP} 
with $\alpha_{\mathrm{HF}}=0.53$. 
The DFT and MP2 calculations are performed with the 6-311++G(3df,3pd) basis set, where the reference values of binding energy are obtained from Ref.\ \cite{Schaefer}. 
For efficiency, the resolution-of-identity (RI) approximation \cite{KF1997} is used for calculations with the MP2 correlation (using sufficiently large auxiliary basis sets). 

The CCSD(T) dissociation curves are calculated on the fixed monomer geometries (using the CCSD(T) optimized geometry of Ref.\ \cite{Kim}), with the aug-cc-pVTZ basis set. 
Note that although the ZPE corrected energy of the proton transferred structure (or, referred to as the Ion structure in Ref.\ \cite{Kim}.) is inconsistent with Ref.\ \cite{Schaefer}. However, adopting the geometries obtained from Ref.\ \cite{Kim} yields results that are consistent with Ref.\ \cite{Schaefer}. 

All of the calculations are performed with the development version of \textsf{Q-Chem 3.2} \cite{QChem3}. As the basis set superposition error (BSSE) for the ionized water dimer has been shown to be 
insignificant (if the diffuse basis functions are adopted) \cite{JPCA1999,Kim}, we do not perform BSSE correction throughout this paper.

\section{Results and Discussion}

The ZPE corrected binding energies and relative energies of the water dimer cation calculated by various XC functionals are shown in Table \ref{Table:binding} and Table \ref{Table:relative}, respectively. 
The calculated dissociation curves for the hemibonded structure are shown in Fig.~\ref{Fig:DOO}. A summary of the results based on these three different criteria is shown in Table \ref{Table:conclusion}. The 
notation used for characterizing statistical errors is as follows: Mean signed errors (MSEs), mean absolute errors (MAEs) and root-mean-square (RMS) errors. 

\subsection{Criterion I: Binding Energies}

Table~\ref{Table:binding} shows the binding energies of the two structures of water dimer radical cation and the transition state between them. The reference data obtained from Ref.\ \cite{Schaefer} are 
based on the CCSD(T) calculations with the aug-cc-pVQZ basis set. Since the errors of XC functionals for the hemibonded structure are much larger than those for the proton transferred structure, we focus our discussion on 
the hemibonded structure. 
From Table \ref{Table:binding}, the functionals with MAE less than 2.5 kcal/mol are $\omega\mbox{B97X-2(LP)}$, M06HF, and BH\&HLYP. Table \ref{Table:binding} also confirms the trend that has been mentioned 
previously: Functionals with larger fractions of HF exchange give more accurate results for the hemibonded structure. To give reasonable results for the hemibonded structure, the $\alpha_{\mathrm{HF}}$ of 
a global hybrid functional should be at least larger than 0.43, as observed in Ref.\ \cite{Kim} for the MPW1K functional. Although M06HF, containing a full HF exchange, gives a small error of the hemibonded structure, 
it yields a large error for the proton transferred structure (due to the incomplete cancelation of errors between the exact exchange and semilocal correlation), as shown in Table \ref{Table:binding}. 
Although functionals with $\alpha_{\mathrm{HF}}$ larger than 0.43 have been suggested,  functionals with $\alpha_{\mathrm{HF}}\geq0.5$ are not always reliable, which can be observed from the 
errors of the hemibonded structure calculated by B2PLYP, M05-2X and M06-2X. But this trend still holds: the results of the M05-2X and M06-2X are much better than those of the M05 and M06. 
This means that although the energy of the hemibonded structure is sensitive to the $\alpha_{\mathrm{HF}}$ values in XC functionals, it may also be affected by the associated density functional approximations (DFAs). 
Also note that some GGA functionals, such as BLYP and PBE, cannot predict that the transition state between the two structures of the water dimer radical cation.

The HF exchange included in the $\omega\mbox{B97}$ series is given by 
\begin{equation}
E^{\omega\mathrm{B97~series} }_{\mathrm{HF~exchange} }=E^{\mathrm{LR-HF} }_x+C_xE^{\mathrm{SR-HF} }_x,
\label{Eq:wB97}
\end{equation}
where
\begin{align}
E^{\mathrm{LR-HF}}_x=&-\frac{1}{2}\sum_{\sigma}\sum^{occu.}_{i,j}\int\int\psi^*_{i\sigma}(\textbf{r}_1)\psi^*_{j\sigma}(\textbf{r}_2)\nonumber\\ 
&\times\frac{\mbox{erf}(\omega r_{12})  }{r_{12}}\psi_{j\sigma}(\textbf{r}_1)\psi_{i\sigma}(\textbf{r}_2)d\textbf{r}_1d\textbf{r}_2,
\label{Eq:LRHF}
\end{align}
and
\begin{align}
E^{\mathrm{SR-HF} }_x=&-\frac{1}{2}\sum_{\sigma}\sum^{occu.}_{i,j}\int\int\psi^*_{i\sigma}(\textbf{r}_1)\psi^*_{j\sigma}(\textbf{r}_2)\nonumber\\
&\times\frac{\mbox{erfc}(\omega r_{12})  }{r_{12}}\psi_{j\sigma}(\textbf{r}_1)\psi_{i\sigma}(\textbf{r}_2)d\textbf{r}_1d\textbf{r}_2.
\label{Eq:SRHF}
\end{align}
Here $r_{12}\equiv\left|{\bf r}_{12}\right|=\left|{\bf r}_{1}-{\bf r}_{2}\right|$ (atomic units are used throughout this paper). The parameter $\omega$ defines the range of the splitting operators. 
The coefficients for the $\omega\mbox{B97}$ series are listed in Table \ref{Table:wB97}. Since the fraction of HF exchange in $\omega\mbox{B97}$ series depends on the interelectronic distance $r_{12}$, the trend 
mentioned previously is not as obvious as the global hybrid functionals. But it is clear that the $\omega\mbox{B97X-2(LP)}$, a LC double-hybrid functional, gives the most accurate results compare to the 
other functionals in the $\omega\mbox{B97}$ series.

\begin{widetext}
\begin{table*}
\begin{threeparttable}
\caption{Binding energies (in kcal/mol) of the ionized water dimer.}
\begin{tabular}{@{}lcccccccccc@{} }
\hline\hline
  & & \multicolumn{2}{p{4.5cm}}{Proton transferred structure} & \multicolumn{2}{p{3cm}}{Transition state} & \multicolumn{2}{p{4.5cm}}{Hemibonded structure}  & & & \\
\cline{3-8}
Method   &   $\alpha_{\mathrm{HF}}$      &   E   & Error     &  E    &  Error      &   E   &   Error   &   MSE   &   MAE   &   RMS \\
\hline
BLYP & 0.00 & -45.62 & 2.10 & - & - & -52.89 & 18.17 & - & - & -\\

PBE & 0.00 & -47.37 & 3.85 & - & - & -53.93 & 19.21 & - & - & -\\

M06L & 0.00 & -45.00 & 1.48 & -40.85 & 12.45 & -48.24 & 13.52 & 9.15 & 9.15 & 10.65\\ 

B97 & 0.19 & -45.33 & 1.81 & -38.39 & 9.99 & -45.88 & 11.16 & 7.65 & 7.65 & 8.71\\

B3LYP & 0.20 & -45.94 & 2.42 & -38.43 & 10.02 & -45.67 & 10.95 & 7.80 & 7.80 & 8.68\\

PBE0 & 0.25 & -46.61 & 3.09 & -36.75 & 8.35 & -43.95 & 9.23 & 6.89 & 6.89 & 7.40\\

M06 & 0.27 & -45.83 & 2.31 & -36.13 & 7.73 & -42.43 & 7.71 & 5.92 & 5.92 & 6.44\\

M05 & 0.28 & -45.35 & 1.83 & -35.46 & 7.06 & -41.31 & 6.59 & 5.16 & 5.16 & 5.68\\

BH\&HLYP & 0.50 & -45.97 & 2.45 & -29.55 & 1.15 & -35.11 & 0.39 & 1.33 & 1.33 & 1.58\\

B2PLYP & 0.53 & -45.04 & 1.52 & -32.58 & 4.18 & -40.96 & 6.24 & 3.98 & 3.98 & 4.42\\

M06-2X & 0.54 & -47.05 & 3.53 & -32.21 & 3.81 & -40.13 & 5.41 & 4.25 & 4.25 & 4.33\\ 

M05-2X & 0.56 & -46.76 & 3.24 & -31.99 & 3.59 & -39.35 & 4.63 & 3.82 & 3.82 & 3.87\\

$\omega$B97 & 0.00---1.00 & -45.92 & 2.40 & -33.63 & 5.23 & -41.66 & 6.94 & 4.86 & 4.86 & 5.20\\

$\omega$B97X & 0.16---1.00 & -46.13 & 2.61 & -34.34 & 5.94 & -42.20 & 7.48 & 5.34 & 5.34 & 5.72\\

$\omega$B97X-D & 0.22---1.00 & -45.71 & 2.19 & -35.33 & 6.93 & -43.14 & 8.42 & 5.85 & 5.85 & 6.42\\

$\omega$B97X-2(LP) & 0.68---1.00 & -45.42 & 1.90 & -29.17 & 0.77 & -37.83 & 3.11 & 1.93 & 1.93 & 2.15\\

M06HF & 1.00 & -48.36 & 4.84 & -28.23 & -0.17 & -35.75 & 1.03 & 1.90 & 2.01 & 2.86\\

MP2 & 1.00 & -43.95 & 0.43 & -25.16 & -3.24 & -30.03 & -4.69 & -2.50 & 2.79 & 3.30\\

CCSD(T)\tnote{a} & 1.00 & -43.52\tnote{b} & 0.00 & -28.39 & 0.00 & -34.72 & 0.00 & 0.00 & 0.00 & 0.00\\
\hline\hline
\end{tabular}
\label{Table:binding} 
\begin{tablenotes}
\item[a] The CCSD(T) results, taken from Ref.\ \cite{Schaefer}, is adopted as the reference. 
\item[b] The ZPE corrected binding energy of the proton transferred structure calculated by CCSD(T) in Ref.\ \cite{Kim} is inconsistent with Ref.\ \cite{Schaefer}. However, adopting the geometry of Ref.\ \cite{Schaefer} will yield the results that are consistent with Ref.\ \cite{Schaefer}.
\end{tablenotes}
\end{threeparttable}

\end{table*}
\end{widetext}

As mentioned previously, functionals with large fractions of HF exchange may perform well for the hemibonded structure where the serious SIE takes place, they may perform unsatisfactorily for the other 
structures. Therefore, we also consider another criterion: the relative energies between the three structures, as proposed by Cheng \textit{et al.} \cite{Schaefer}.

\subsection{Criterion II: Relative energies}

In this criterion, the ground-state energy of the proton transferred structure is set to the zero point, \textit{i.e.} both of the ground state energies of the hemibonded structure and that of the transition state are relative to 
the proton transferred structure, as shown in Table \ref{Table:relative}. The previously recommended functional, M06HF, performs poorly here. 

In this criterion, it is obvious that functionals without the exact HF exchange, such as BLYP and PBE, overstabilize the hemibonded structure and wrongly predict the hemibonded structure to be more stable than 
the proton transferred one. 

In addition to the previously recommended functionals based on Criterion I, the M05-2X and the M06-2X functionals also give accurate relative energies here. Although they give results that are not accurate enough for the 
binding energies, they yield good relative energies between those three structures of the water dimer radical cation.

In fact, functionals that are unable to give reasonable binding energies for the hemibonded structure may be traced back to the predicted dissociation curves of the hemibonded systems. 
Since many functionals fail to dissociate the hemibonded structure of water dimer radical cation into the correct fragments. The definition of the binding energy is not well-defined. 
Therefore, the entire dissociation curve from the hemibonded structure should be concerned. 

\begin{widetext}
\begin{table*}
\begin{threeparttable}
\caption{Relative energies (in kcal/mol) between those three structures of the water dimer radical cation.}
\begin{tabular}{@{}lcccccccccc@{} }
\hline\hline
  & & \multicolumn{2}{p{4.5cm}}{Proton transferred structure} & \multicolumn{2}{p{3cm}}{Transition state} & \multicolumn{2}{p{4.5cm}}{Hemibonded structure}  & & & \\
\cline{3-8}
Method   &   $\alpha_{\mathrm{HF}}$      &   E   & Error     &  E    &  Error      &   E   &   Error   &   MSE   &   MAE   &   RMS \\
\hline
BLYP & 0.00 & 0.00 & 0.00 & - & - & -7.27 & 16.07 & - & - & -\\
 
PBE & 0.00 & 0.00 & 0.00 & - & - & -6.56 & 15.36 & - & - & -\\ 

M06L & 0.00 & 0.00 & 0.00 & 4.14 & 10.86 & -3.24 & 12.04 & 11.45 & 11.45 & 11.47\\ 

B97 & 0.19 & 0.00 & 0.00 & 6.93 & 8.07 & -0.55 & 9.35 & 8.71 & 8.71 & 8.74\\ 

B3LYP & 0.20 & 0.00 & 0.00 & 7.51 & 7.49 & 0.27 & 8.53 & 8.01 & 8.01 & 8.02\\ 

PBE0 & 0.25 & 0.00 & 0.00 & 9.86 & 5.14 & 2.66 & 6.14 & 5.64 & 5.64 & 5.66\\ 

M06 & 0.27 & 0.00 & 0.00 & 9.70 & 5.30 & 3.40 & 5.40 & 5.35 & 5.35 & 5.35\\ 

M05 & 0.28 & 0.00 & 0.00 & 9.89 & 5.11 & 4.04 & 4.76 & 4.94 & 4.94 & 4.94\\ 

BH\&HLYP & 0.50 & 0.00 & 0.00 & 16.41 & -1.41 & 10.86 & -2.06 & -1.74 & 1.74 & 1.77\\ 

B2PLYP & 0.53 & 0.00 & 0.00 & 12.46 & 2.54 & 4.08 & 4.72 & 3.63 & 3.63 & 3.79\\ 

M06-2X & 0.54 & 0.00 & 0.00 & 14.85 & 0.15 & 6.92 & 1.88 & 1.02 & 1.02 & 1.33\\ 

M05-2X & 0.56 & 0.00 & 0.00 & 14.77 & 0.23 & 7.41 & 1.39 & 0.81 & 0.81 & 1.00\\ 

$\omega$B97 & 0.00---1.00 & 0.00 & 0.00 & 12.29 & 2.71 & 4.27 & 4.53 & 3.62 & 3.62 & 3.73\\ 

$\omega$B97X & 0.16---1.00 & 0.00 & 0.00 & 11.79 & 3.21 & 3.92 & 4.88 & 4.05 & 4.05 & 4.13\\ 

$\omega$B97X-D & 0.22---1.00 & 0.00 & 0.00 & 10.38 & 4.62 & 2.57 & 6.23 & 5.42 & 5.42 & 5.48\\ 

$\omega$B97X-2(LP) & 0.68---1.00 & 0.00 & 0.00 & 16.25 & -1.25 & 7.59 & 1.21 & -0.02 & 1.23 & 1.23\\ 

M06HF & 1.00 & 0.00 & 0.00 & 20.12 & -5.13 & 12.61 & -3.81 & -4.47 & 4.47 & 4.52\\ 

MP2 & 1.00 & 0.00 & 0.00 & 18.79 & -3.79 & 13.92 & -5.12 & -4.45 & 4.45 & 4.50\\ 

CCSD(T)\tnote{a} & 1.00 & 0.00 & 0.00 & 15.14 & 0.00 & 8.80 & 0.00 & 0.00 & 0.00 & 0.00\\ 
\hline\hline
\end{tabular}
\label{Table:relative}
\begin{tablenotes}
\item[a] The CCSD(T) results, taken from Ref.\ \cite{Schaefer}, is adopted as the reference. 
\end{tablenotes}
\end{threeparttable}
\end{table*}
\end{widetext}

\begin{table}[h]
\caption{The coefficients of the SR HF exchange and $\omega$ for $\omega$B97 series.}
\begin{tabular}{ p{1.5cm}p{1.5cm}p{1.5cm}p{1.5cm}p{2cm} }
\hline\hline
　                              &    $\omega$B97 & $\omega$B97X & $\omega$B97X-D & $\omega$B97X-2(LP)\\ 

\hline

$\omega$(bohr$^{-1}$)  &  0.4         & 0.3         & 0.2            & 0.3\\ 
$C_x$                       &    0.00         & 0.16     & 0.22        & 0.68\\ 
\hline\hline
\end{tabular}
\label{Table:wB97}
\end{table}

\subsection{Criterion III: Dissociation behavior\label{subsect:Disso}}

\begin{figure}[h]
(a)
\includegraphics[scale=0.3]{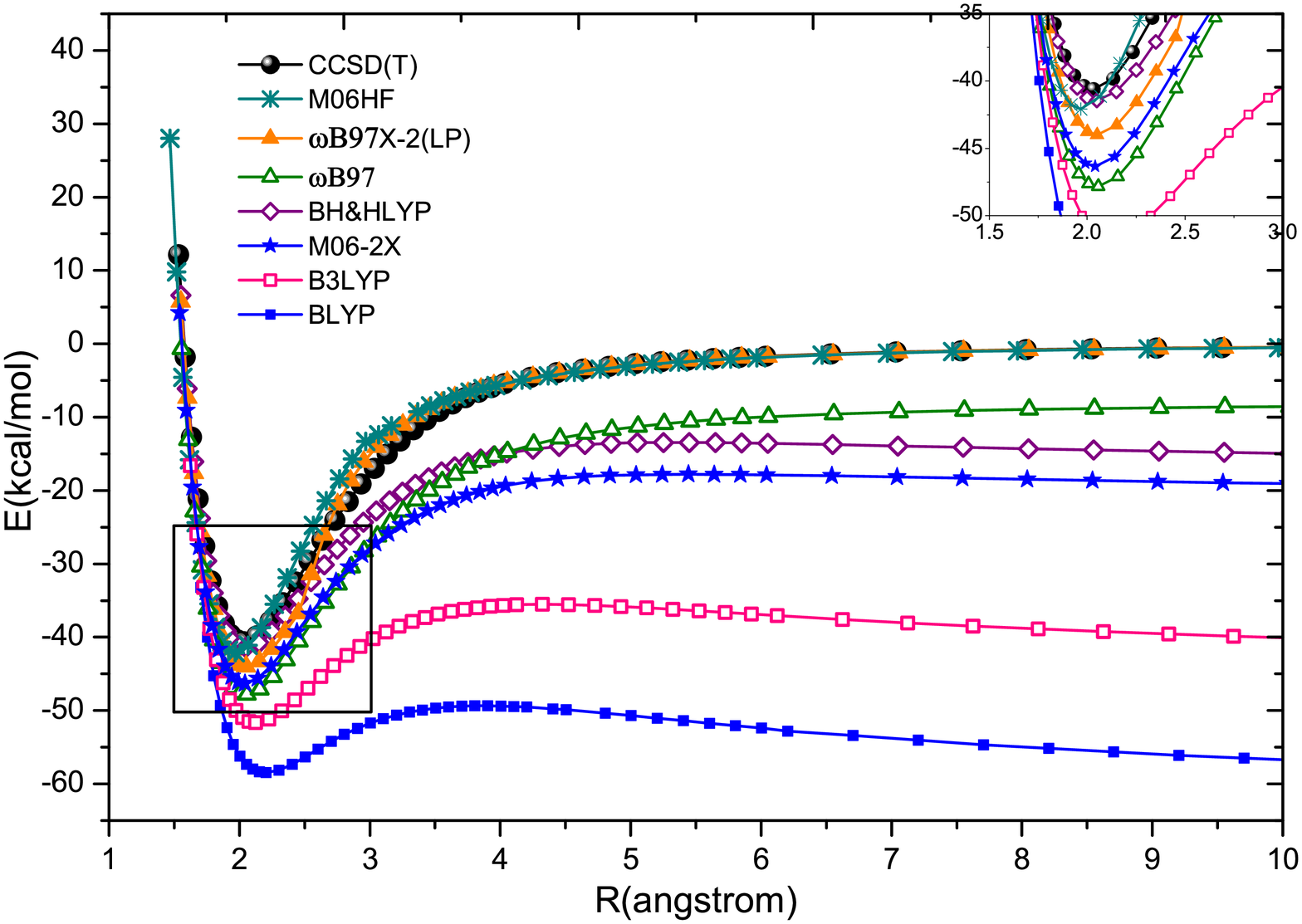}
(b)
\includegraphics[scale=0.3]{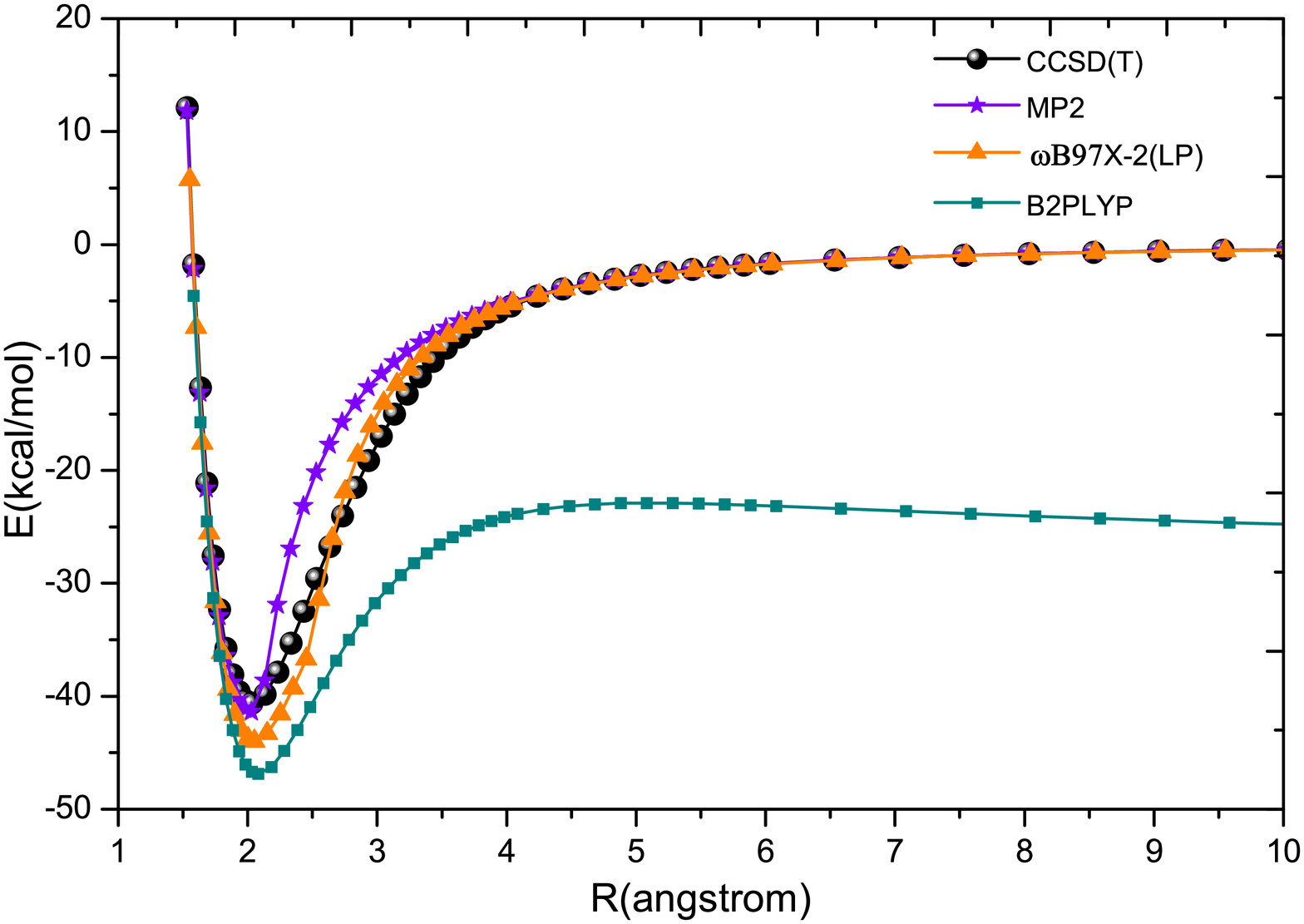}
\caption{(a) Dissociation curves for the hemibonded structure calculated by various XC functionals. (b) Dissociation curves for the hemibonded structure calculated by MP2 and double-hybrid functionals.}
\label{Fig:DOO}
\end{figure}

Due to the severe SIEs associated with DFAs, systems with three-electron hemibonds, such as the hemibonded structure of water dimer radical cation, are especially difficult for conventional density functionals. 
Many XC functional cannot dissociate it into the correct fragments, H$_2$O and H$_2$O$^+$ (ionic state), \textit{i.e.}~they predict that the hemibonded structure should be dissociated into two fragments, each of which 
carries half of positive charge (covalent state). Fig.~\ref{Fig:DOO} shows the dissociation curves calculated by various XC functionals. Note that the discontinuous points in Fig.~\ref{Fig:DOO} near $R$ = 2.5 angstrom 
for the $\omega$B97X-2(LP) and 3 angstrom for the M06HF functional are the respective broken-symmetry points. 

The BH\&HLYP functional, which has been suggested in the equilibrium ground-state energy calculation by the earlier reports \cite{JPCA1999,Kim} has a spurious barrier on its dissociation curve. 
Although we do not present the dissociation curve of the MPW1K functional, we expect it will suffer from the same problem as BH\&HLYP. 

The spurious barrier can be removed if the 100\% exact exchange is adopted in a functional (e.g.\ M06HF), but its shortcoming is described in the previous subsection and is thus not recommended. This shortcoming can 
be greatly reduced by the use of LC hybrid functionals, such as the $\omega$B97 series or the other LC hybrid functionals \cite{wM05-D}. The LC functionals retain the full HF exchange at the long range, while the good 
cancelation of errors between the semilocal exchange and correlation functionals are retained at the short range \cite{omegaB97}. 

\begin{widetext}
\begin{table*}[h]
\begin{threeparttable}
\caption{Summary of results based on the three criteria.}
\begin{tabular}{@{}lcccccccc@{} }
\hline\hline
 & 　 & \multicolumn{7}{p{12cm}}{Criteria} \\
 \cline{3-9}
 &  & \multicolumn{3}{p{3cm}}{Binding Energies} & \multicolumn{3}{p{3cm}}{Relative Energies} & Correct dissociation limit\\
  \cline{3-9}
Method & $\alpha_{\mathrm{HF}}$ & MSE & MAE & RMS & MSE & MAE & RMS & \\
 \hline
BLYP & 0.00 & - & - & - & - & - & - & No\\

PBE & 0.00 & - & - & - & - & - & - & No\\

M06L & 0.00 & 9.15 & 9.15 & 10.65 & 11.45 & 11.45 & 11.47 & No\\

B97 & 0.19 & 7.65 & 7.65 & 8.71 & 8.71 & 8.71 & 8.74 & No\\

B3LYP & 0.20 & 7.80 & 7.80 & 8.68 & 8.01 & 8.01 & 8.02 & No\\

PBE0 & 0.25 & 6.89 & 6.89 & 7.40 & 5.64 & 5.64 & 5.66 & No\\

M06 & 0.27 & 5.92 & 5.92 & 6.44 & 5.35 & 5.35 & 5.35 & No\\

M05 & 0.28 & 5.16 & 5.16 & 5.68 & 4.94 & 4.94 & 4.94 & No\\

BH\&HLYP & 0.50 & 1.33 & 1.33 & 1.58 & -1.74 & 1.74 & 1.77 & No\\

B2PLYP & 0.53 & 3.98 & 3.98 & 4.42 & 3.63 & 3.63 & 3.79 & No\\
 
M06-2X & 0.54 & 4.25 & 4.25 & 4.33 & 1.02 & 1.02 & 1.33 & No\\

M05-2X & 0.56 & 3.82 & 3.82 & 3.87 & 0.81 & 0.81 & 1.00 & No\\

$\omega$B97 & 0.00---1.00 & 4.86 & 4.86 & 5.20 & 3.62 & 3.62 & 3.73 & No\\
 
$\omega$B97X & 0.16---1.00 & 5.34 & 5.34 & 5.72 & 4.05 & 4.05 & 4.13 & No\\
 
$\omega$B97X-D & 0.22---1.00 & 5.85 & 5.85 & 6.42 & 5.42 & 5.42 & 5.48 & No\\
 
$\omega$B97X-2(LP) & 0.68---1.00 & 1.93 & 1.93 & 2.15 & -0.02 & 1.23 & 1.23 & Yes\\

M06HF & 1.00 & 1.90 & 2.01 & 2.86 & -4.47 & 4.47 & 4.52 & Yes\\

MP2 & 1.00 & -2.50 & 2.79 & 3.30 & -4.45 & 4.45 & 4.50 & Yes\\
 
CCSD(T)\tnote{a} & 1.00 & 0.00 & 0.00 & 0.00 & 0.00 & 0.00 & 0.00 & Yes\\ 
\hline\hline
\end{tabular}
\label{Table:conclusion} 
\begin{tablenotes}
\item[a] The CCSD(T) results, taken from Ref.\ \cite{Schaefer}, is adopted as the reference. 
\end{tablenotes}
\end{threeparttable}

\end{table*}
\end{widetext}

In the following, we will explain why LC hybrid functionals do not suffer from a spurious energy barrier as global hybrid functionals do. An estimate of the SIE of symmetric radical cations by global hybrid or pure density functionals has been derived as: \cite{PCCP, Grafen2004}
\begin{equation}
E^{\mathrm{SIE} }\approx(1-\alpha_{ \mathrm{HF} } )\left[\left(\frac{1}{2}-C\right)J+\frac{1}{4R}\right],
\label{Eq:SIE}
\end{equation}
where $C\approx2^{-1/3}$, $(0.5-C)\approx-0.29$, and $J$ is the Coulomb self-interaction energy for the ionic state (in the case that the bond electron is localized at either of the two fragments). For three-electron-bonded radical 
cations, the bonding between the fragments is accomplished by the delocalized $\beta$ electron, which dominates the total SIE \cite{PCCP}. 

We have derived an estimate of the SIE of symmetric radical cations by LC hybrid functionals, with the details arranged in the appendix. And the result is
\begin{equation}
E^{\mathrm{SIE} }\approx(1-C_x)\left[\left(\frac{1}{2}-B(\omega)C\right)J^\text{SR}(\omega)+\frac{\text{erfc}(\omega R)}{4R}\right],
\label{LC_SIE}
\end{equation}
where $B(\omega)$ and $J^\text{SR}(\omega)$ are constants with respect to $R$. But they depend on $\omega$, \textit{i.e.}~for LC hybrid functionals with different $\omega$ values, their $B(\omega)$ and $J^\text{SR}(\omega)$ are different. The dependence of $J^\text{SR}(\omega)$ on $\omega$ is defined by
\begin{equation}
J^\mathrm{SR}(\omega)=\frac{1}{2}\iint \rho_{\beta}(\textbf{r}_1)\frac{\text{erfc}(\omega r_{12})}{r_{12}}\rho_{\beta}(\textbf{r}_2)d\textbf{r}_1d\textbf{r}_2,
\end{equation}
and for small $\omega$,
\begin{equation}
B(\omega)\approx1-0.254\omega(\text{bohr}^{-1}).
\label{B_approx}
\end{equation}

\begin{figure}[h]
(a)
\includegraphics[scale=0.35]{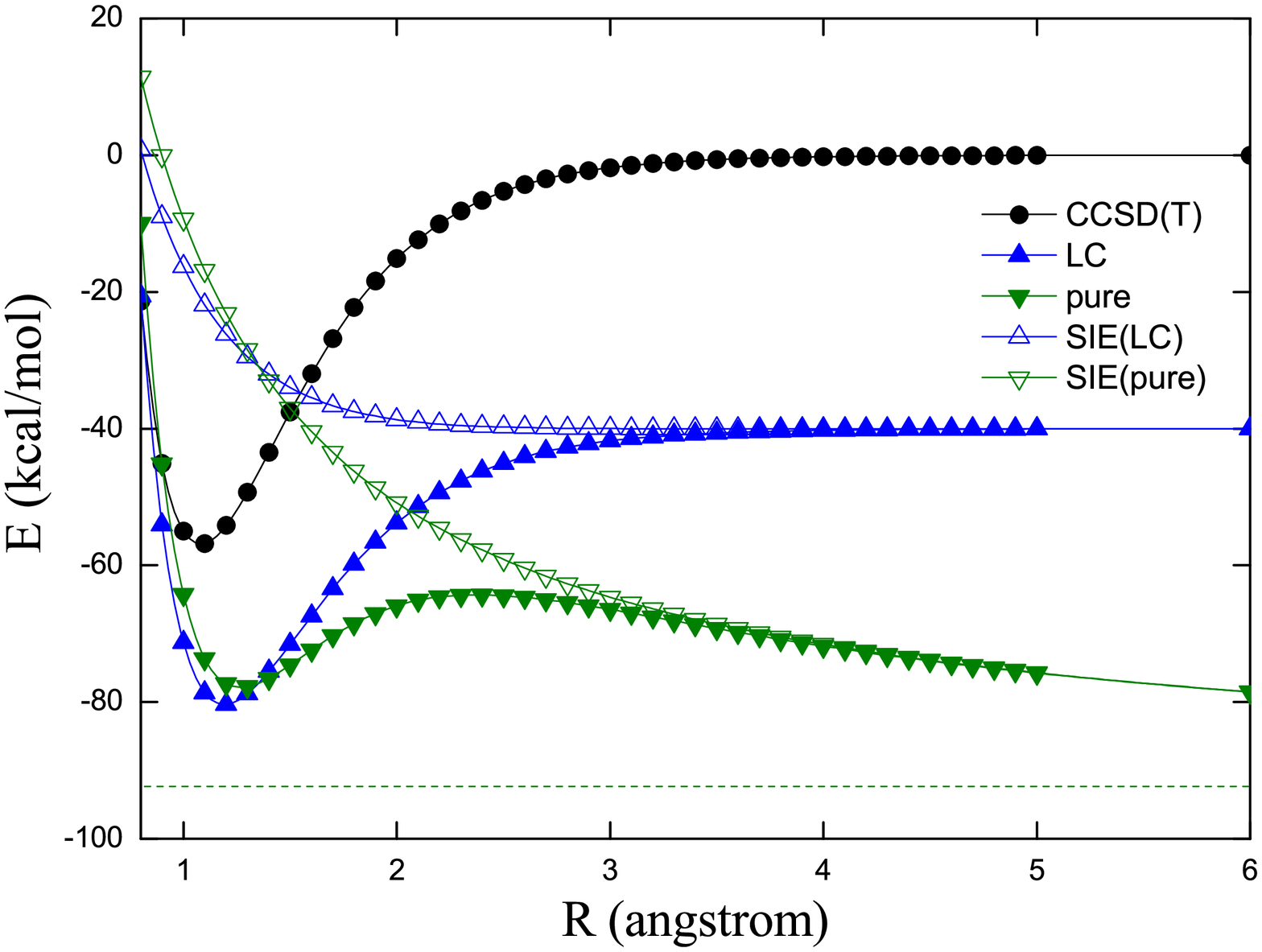} 
(b)
\includegraphics[scale=0.35]{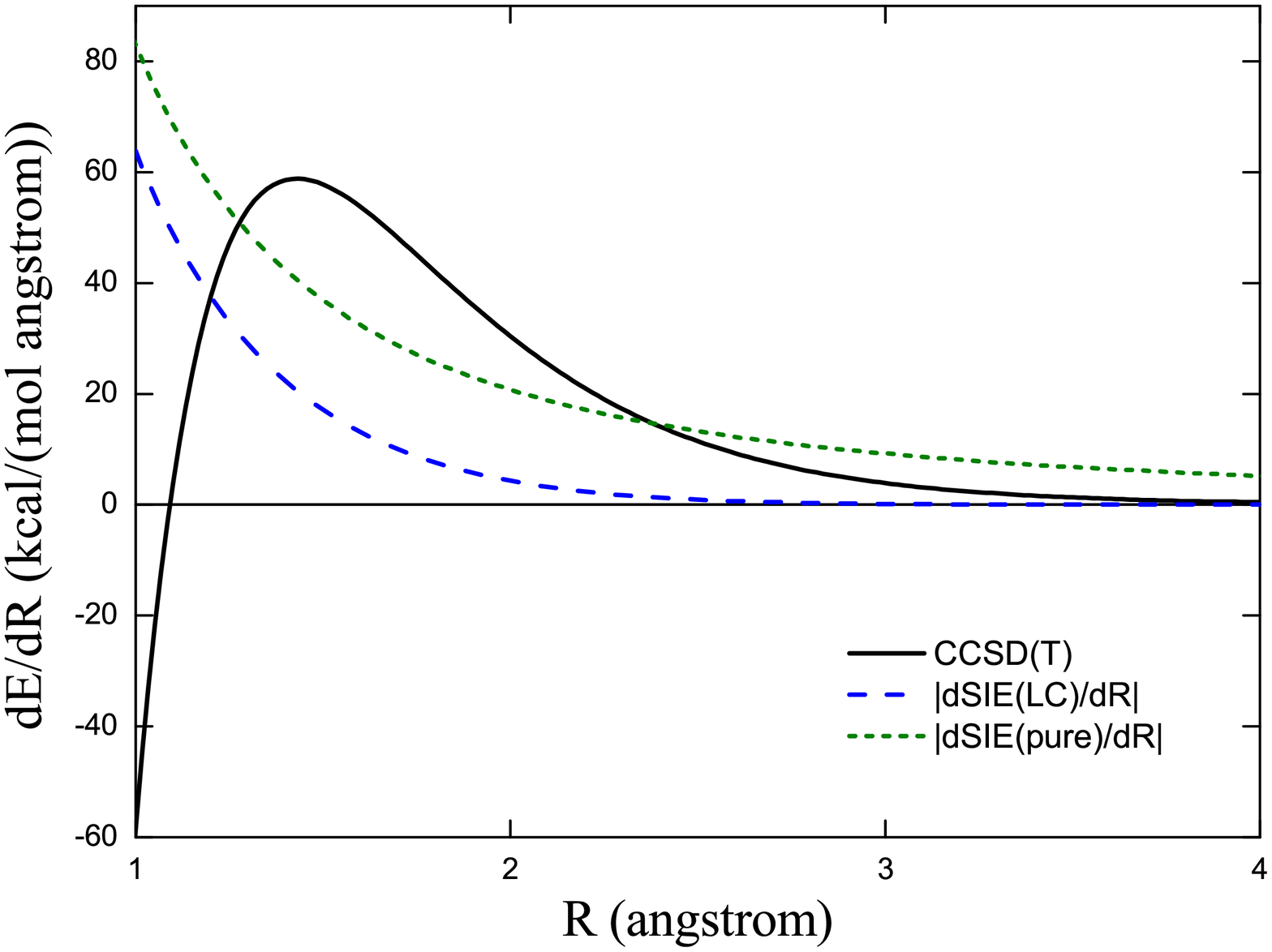} 
\caption{The spurious energy barrier of the hemibonded systems predicted by DFT functionals can be illustrated qualitatively. (a) Comparison of \ce{He2+} dissociation curves by pure DFT and LC hybrid functional using 
estimate~(\ref{LC_SIE}). Zero level is set to E(He)+E(\ce{He2+}) for each method. (b) First derivative of exact \ce{He2+} dissociation curve and SIE derivative magnitude curves by pure DFT and LC hybrid functional 
using estimate~(\ref{dEdR_approx}).}
\label{Fig:SIE}
\end{figure}

Note that estimate~(\ref{Eq:SIE}) is the special case with vanishing $\omega$ of estimate~(\ref{LC_SIE}). We apply estimate~(\ref{LC_SIE}) and (\ref{B_approx}) to the simplest three-electron-bonded 
system, and the estimated \ce{He2+} (which is also a three-electron hemibonded system) dissociation curves for pure DFT and LC hybrid functional are shown in Fig.~\ref{Fig:SIE}(a). For simplicity, $C_x$ is set to zero 
and the LDA orbital is used for evaluating the Coulomb self-interactions; $\omega$ is set to 0.4 bohr$^{-1}$ for the LC hybrid functional. When the exact dissociation curve approaches zero, the SIE of the LC hybrid functional 
is already close to a constant, while the SIE of pure DFT is still decreasing. Therefore the dissociation curve by the LC hybrid functional does not display a spurious energy barrier as that of the pure DFT. Another effect 
of the long-range correction is the reduction of Coulomb self-interaction for the ionic state. In Fig.~\ref{Fig:SIE}(a), $(0.5-C)J\approx-92$ kcal/mol has been modified to $[0.5-B(\omega)C]J^\text{SR}(\omega)\approx-40$ kcal/mol.

The formal feature of spurious energy barrier is one more turning point (at which the derivative changes sign) on top of the barrier, in addition to the one in the potential well. Since the ground state energy calculated by a functional is approximately the exact ground state energy plus the SIE produced by that functional,
\begin{equation}
E^{\mathrm{DFT}}\approx{}E^{ \mathrm{exact} }+E^{ \mathrm{SIE} },
\label{Eq:Exact+SIE}
\end{equation}
turning points occur when 
$|dE^\text{SIE}/dR|$ equals the derivative of the exact curve, \textit{i.e.}~points where $|dE^\text{SIE}/dR|$ intersects the derivative of the exact curve. $|dE^\text{SIE}/dR|$ by pure DFT is $1/4R^2$. For the LC hybrid functional, there is an extra multiplicative factor:
\begin{equation}
4R^2\left\vert\frac{dE^\text{SIE}}{dR}\right\vert\approx\frac{2\omega R}{\sqrt{\pi}}\text{exp}\left[-(\omega R)^2\right]+\text{erfc}(\omega R).
\label{dEdR_approx}
\end{equation}
As shown in Fig.~\ref{Fig:SIE}(b), for sufficient $\omega$ value (0.2 bohr$^{-1}$ or more), this factor can change the decay nature of the derivative magnitude, from power law to exponential. Thus, the SIE derivative 
magnitude curve of typical LC hybrid functionals can avoid second intersection with the derivative of the exact curve. Global hybrid functionals simply scale down the SIE derivative curve by a constant, so they cannot avoid 
the second intersection, unless $\alpha_\text{HF}$ approaches unity. This explains why global hybrid functionals display a spurious energy barrier which LC hybrid functionals avoid.

Note that LC hybrid functionals which do not contain SR HF exchange, such as $\omega$B97, still can be free from the spurious barrier, but will lose the possibility to predict symmetry breaking during the 
dissociation, $i.e.$ the dissociation curve can not converge to zero. This means that the SR HF exchange is also important. In fact, the covalent (symmetric) state and the ionic (symmetry-broken) state are nearly degenerate 
by CCSD(T) calculations \cite{PCCP}. But most of the XC functionals cannot predict that these two states are degenerate: Due to the serious SIE, they usually overstabilize the covalent state. Therefore, functional which 
can predict that the ionic state is more stable than the covalent state (\textit{i.e.}~the hemibonded structure will dissociate into H$_2$O and H$_2$O$^+$) will give correct dissociation limit. Very recently, a double-hybrid 
functional containing a very large fraction of HF exchange ($\approx 79 \%$) \cite{PBE0-2}, has been shown to be promising for reducing the SIEs in hemibonded systems. 

The dissociation curves of the hemibonded structure calculated by double-hybrid functionals and MP2 are shown in Fig.~\ref{Fig:DOO}(b). Note that the PT2 calculation should be executed in the stable wave 
function. For example, although the dissociation curve of the covalent state seems more stable than that of the ionic state, we should choose the dissociation curve of the ionic state as an actual dissociation behavior 
calculated by MP2. Since the HF theory, which provides the reference orbitals for computing the MP2 correlation energy, predicts the ionic state to be more stable than the covalent one. Thus we choose the dissociation 
curve of the ionic state rather than that of the covalent state. Fig.~\ref{Fig:DOO}(b) shows that the $\omega$B97X-2(LP) functional and the MP2 theory can predict the correct dissociation limits. While the B2PLYP 
functional cannot. 

\begin{figure}[h]
\includegraphics[scale=0.3]{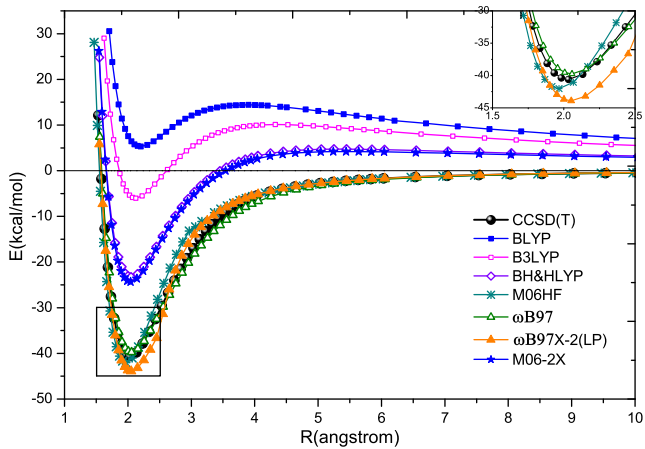} 
\caption{Dissociation curves for the hemibonded structure. The zeros of dissociation curves are set to their respective dissociation limits.}
\label{Fig:Dshift}
\end{figure}

There is another way to define the dissociation behavior of the XC functionals: Since many of the XC functionals cannot predict that the hemibonded structure of the ionized water dimer will dissociate into H$_2$O and 
H$_2$O$^+$, we set the zeros of dissociation curves to their respective dissociation limits, as shown in Fig.~\ref{Fig:Dshift}. In this definition, we focus on the potential curve experienced by the two fragments 
during the dissociation process of the hemibonded structure. Surprisingly, the dissociation curve of the $\omega$B97 functional is extremely close to that of the CCSD(T) theory. This means that the $\omega$B97 gives 
the best potential energy curve toward the dissociation process. The previous suggested functionals \cite{JPCA1999,Kim}, 
such as BH\&HLYP, yield potential curves that are too shallow. Functionals which predict symmetry-breaking solutions during the dissociation process (e.g.\ M06HF and $\omega$B97X-2(LP)) are found to yield 
the dissociation curves that are narrower than that of the CCSD(T) theory. 

\begin{figure}[h]
\includegraphics[scale=0.3]{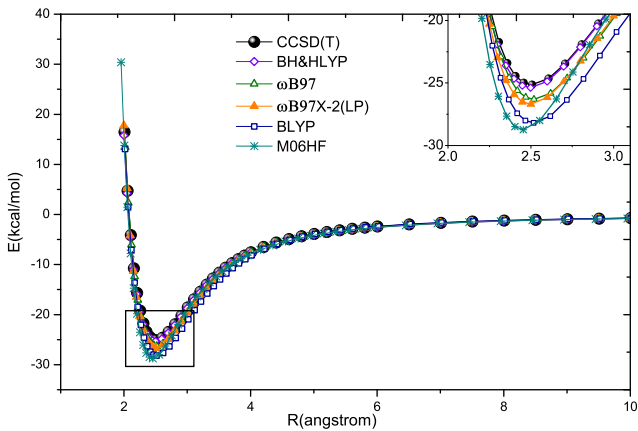} 
\caption{The dissociation curves for the proton transferred structure of ionized water dimer.}
\label{Fig:Dproton}
\end{figure}

Finally, we discuss the dissociation of the proton transferred structure of the water dimer cation. In this structure, the SIEs associated with functionals for this structure are not as large as those for the hemibonded 
structure, so all the dissociation curves are very similar, as shown in Fig.~\ref{Fig:Dproton}.

The results of the water dimer radical cation using three criteria we proposed are summarized in Table.~\ref{Table:conclusion}. We find that the functional which performs well based on these three criteria is the 
$\omega$B97X-2(LP) functional, yielding the accurate binding energies, relative energies, and the correct dissociation limit. However, this functional yields a dissociation curve of the hemibonded structure that is 
a little narrower than that of CCSD(T). For applications sensitive to the shape of potential of the hemibonded structure, we suggest to use $\omega$B97: Although this functional neither gives a 
dissociation curve of the hemibonded structure that converges to zero nor yields accurate binding energies, this functional gives a dissociation curve which has nearly the same shape as the curve calculated by 
CCSD(T). Thus, we recommend this functional for researchers who like to perform the molecular dynamics of the water dimer cation. 

\section{Conclusions}

We have proposed three criteria to examine the performance of density functionals on water dimer radical cation, and explained why LC hybrid functionals generally work better than conventional density functionals 
for hemibonded systems. 

The previously recommended functional, BH\&HLYP, cannot dissociate the hemibonded structure of the water dimer cation into the correct fragments: H$_2$O and H$_2$O$^+$. Furthermore, the BH\&HLYP dissociation curve 
displays an unphysical repulsive barrier, and is too shallow for molecular dynamics simulations. Such a spurious barrier could be removed by functionals with very large fractions of HF exchange (e.g.\ M06HF 
and $\omega$B97X-2(LP)). LC hybrid functionals, such as the $\omega$B97 series, are shown to be accurate for dissociation curves of the hemibonded structure (\textit{i.e.}~no spurious barriers), and thus are suitable for 
molecular dynamics simulations of larger-size systems. 
For research which is sensitive to the dissociation curve experienced by the fragments of the hemibonded structure, we recommend the use of the $\omega$B97 functional. For researchers who do not sure which criterion is the most important factor during the simulation of water dimer radical cation, we recommend the  use of the $\omega$B97X-2(LP) functional, which is overall  good throughout those three criteria we proposed.

\appendix
\renewcommand{\theequation}{A\arabic{equation}}
\setcounter{equation}{0}  
\section*{Appendix}

We first reproduce estimate~(\ref{Eq:SIE}) for the SIE of symmetric radical cations by global hybrid or pure density functionals \cite{PCCP,Grafen2004}.  Since most XC functionals predict that the covalent state is more stable than the ionic state, the SIE of the covalent state is of interest. Neglecting the small SIE in the correlation energy, we have
\begin{equation}
E^{\mathrm{SIE} }_{\mathrm{cov} }\approx J^{ \mathrm{SI} }_{ \mathrm{cov} }+E^{ \mathrm{SI} }_{ \mathrm{x,~cov} }.
\label{Eq:Ecov}
\end{equation}
Because the SIE is small in the ionic state,
\begin{equation}
E^{ \mathrm{SI} }_{ \mathrm{x,~ionic} }\approx{}-J^{ \mathrm{SI} }_{ \mathrm{ionic} },
\end{equation}
it is favorable to express $J^{ \mathrm{SI} }_{ \mathrm{cov} }$ and $E^{ \mathrm{SI} }_{ \mathrm{x,~cov} }$ in terms of the Coulomb self-interaction energy for the ionic state $J^{ \mathrm{SI} }_{ \mathrm{ionic} }$, or simply $J$ in subsection \ref{subsect:Disso}. This can be done by substituting the density of the delocalized $\beta$ electron (which causes the serious SIE of the hemibonding systems \cite{PCCP}) in the covalent state with the density of that electron in the ionic state. Note that there are two situations for the ionic state: one is the electric charge localized in fragment A and the other is the electric charge localized in fragment B. To a good approximation the density of the delocalized $\beta$ electron of the covalent state can be expressed as
\begin{equation}
\rho^{\beta}_{\mathrm{cov} }(\textbf{r})\approx\frac{ \rho^{\beta}_{\mathrm{A} }(\textbf{r}) }{2}+\frac{ \rho^{\beta}_{\mathrm{B} }(\textbf{r}) }{2}.
\label{Eq:rho}
\end{equation}
The Coulomb self-interaction for the covalent state can be expressed as
\begin{equation}
J^{ \mathrm{SI} }_{ \mathrm{cov} }\approx\frac{1}{2}J^{ \mathrm{SI} }_{ \mathrm{ionic} }+\frac{1}{4R},
\label{Jcov}
\end{equation}
if one assumes that $R$ is large compared to the spatial extent of $\rho^{\beta}_\mathrm{A}$ and $\rho^{\beta}_\mathrm{B}$. This expression can be applied to the self-interaction HF exchange energy for the covalent state,
\begin{equation}
E^{\mathrm{SI,~HF} }_{\mathrm{x,~cov} }=-J^{ \mathrm{SI} }_{ \mathrm{cov} }.
\end{equation}

The pure-DFT self-exchange energy for the covalent state is
\begin{equation}
E^{ \mathrm{SI,~DFT} }_{ \mathrm{x,~cov} }\approx2E_\text{x}(\rho^{\beta}_\mathrm{A}/2)=CE^{ \mathrm{SI} }_{ \mathrm{x, ionic} }.
\end{equation}
For  LDA, $C=2^{-1/3}\approx0.79$. Combining 
\begin{equation}
E^{ \mathrm{SI} }_{ \mathrm{x,~cov} }=\alpha_{ \mathrm{HF} }E^{ \mathrm{SI,~HF} }_{ \mathrm{x,~cov} }+(1-\alpha_{ \mathrm{HF} })E^{ \mathrm{SI,~DFT} }_{ \mathrm{x,~cov} }
\end{equation} and estimate (\ref{Jcov}), one can obtain estimate (\ref{Eq:SIE}).

The SIE estimate for LC hybrid functionals can be derived from the same manner as the above one for global hybrid functionals. Substituting estimate (\ref{Eq:rho}) into the self-interaction LR HF exchange yields
\begin{equation}
E^{\mathrm{SI,~LR-HF} }_{\mathrm{x,~cov} }\approx{}-\left[\frac{1}{2}{     J^{ \mathrm{SI,~LR} }_{ \mathrm{ionic} }     }(\omega)+\frac{\mbox{erf}(\omega R)}{4R}\right],
\label{Eq:SIE-LRHF}
\end{equation}
where we define
\begin{equation}
J^{ \mathrm{SI,~LR} }_{ \mathrm{ionic} }(\omega)=\frac{1}{2}\iint \rho_{\beta}(\textbf{r}_1)\frac{\text{erf}(\omega r_{12})}{r_{12}}\rho_{\beta}(\textbf{r}_2)d\textbf{r}_1d\textbf{r}_2.
\label{Eq:J-LRHF}
\end{equation}
Since the integration of $J^{ \mathrm{SI,~LR} }_{ \mathrm{ionic} }(\omega)$ is only over one fragment, it is independent of $R$. Likewise, the SR HF exchange of the covalent state is
\begin{equation}
E^{\mathrm{SI,~SR-HF} }_{\mathrm{x,~cov} }\approx{}- \left[\frac{1}{2}{ J^{ \mathrm{SI,~SR} }_{ \mathrm{ionic} } }(\omega)+\frac{\mbox{erfc}(\omega R)}{4R}\right],
\label{Eq:SIE-SRHF}
\end{equation}
where we define
\begin{equation}
J^{ \mathrm{SI,~SR} }_{ \mathrm{ionic} }(\omega)=\frac{1}{2}\iint \rho_{\beta}(\textbf{r}_1)\frac{\text{erfc}(\omega r_{12})}{r_{12}}\rho_{\beta}(\textbf{r}_2)d\textbf{r}_1d\textbf{r}_2.
\label{Eq:J-SRHF}
\end{equation}

The SR-DFA self-exchange energy for the covalent state is
\begin{equation}
E^{ \mathrm{SI,~SR-DFA} }_{ \mathrm{x,~cov} }\approx2E_\text{x}(\rho^{\beta}_\mathrm{A}/2)=B(\omega)CE^{ \mathrm{SI} }_{ \mathrm{x, ionic} }.
\end{equation}
For SR-LDA \cite{Gill96,omegaB97},
\begin{equation}
B(\omega)=\int \rho^{4/3}_{\beta}(\textbf{r})F\left(\frac{2^{1/3}\omega}{k_{F \beta}}\right)d\textbf{r}\left/\int \rho^{4/3}_{\beta}(\textbf{r})F\left(\frac{\omega}{k_{F \beta}}\right)d\textbf{r}\right.
\end{equation}
$k_{F \beta}\equiv( 6\pi^2\rho_\beta(\textbf{r}) )^{1/3}$ is the local Fermi wave vector, and the attenuation function is given by
\begin{align}
F(\lambda)&=1-\frac{2\lambda}{3}\left[2\sqrt{\pi}\text{erf}\left(\lambda^{-1}\right)+\cdots\right] \nonumber\\ 
&\approx1-\frac{4\lambda\sqrt{\pi}}{3}~, \text{for small}~\lambda.
\end{align}
 For small $\omega$, and with the density approximated as one electron in a sphere with Bohr radius $a_0$,
\begin{align}
B(\omega)&\approx1-3^{-5/3}4\sqrt[3]{2}(\sqrt[3]{2}-1)\sqrt[6]{\pi}\omega a_0\nonumber\\ 
&\approx1-0.254\omega(\text{bohr}^{-1}).
\end{align}
Combining
\begin{align}
E^{ \mathrm{SI} }_{ \mathrm{x,~cov} }=&E^{ \mathrm{SI,~LR-HF} }_{ \mathrm{x,~cov} }+C_xE^{ \mathrm{SI,~SR-HF} }_{ \mathrm{x,~cov} }\nonumber\\ 
&+(1-C_x)E^{ \mathrm{SI,~SR-DFA} }_{ \mathrm{x,~cov} }
\end{align} and estimate (\ref{Jcov}), one can obtain estimate (\ref{LC_SIE}).

\begin{acknowledgments}

We thank the supports from National Science Council of Taiwan (Grant No.~NSC99-2113-M-003-007-MY2, NSC98-2113-M-001-029-MY3 and NSC98-2112-M-002-023-MY3) and NCTS of Taiwan. We are grateful to the supports from National Taiwan University (Grant No.~99R70304 and 10R80914-1) and Academia Sinica Research Program on NanoScience and Nano Technology

\end{acknowledgments}

\end{document}